\documentclass{emulateapj}
\usepackage{rotating}
\begin{document}
\title{{\em XMM-Newton} Observations of the Extremely Low Accretion Rate Polars SDSSJ155331.12+551614.5 and SDSSJ132411.57+032050.5} 

\author{Paula Szkody\altaffilmark{1},
Lee Homer\altaffilmark{1},
Bing Chen\altaffilmark{2},
Arne Henden\altaffilmark{3},
Gary D. Schmidt\altaffilmark{4},
Scott F. Anderson\altaffilmark{1},
D. W. Hoard\altaffilmark{5},
Wolfgang Voges\altaffilmark{6},
J. Brinkmann\altaffilmark{7}}

\altaffiltext{1}{Department of Astronomy, Box 351580, University of Washington, Seattle, WA 98195} 
\altaffiltext{2}{XMM-Newton Science Operations Centre, ESA/Vilspa, 28080, Madrid, Spain}
\altaffiltext{3}{Universities Space Research Association/US Naval Observatory, Flagstaff Station, P. O. Box 1149,
Flagstaff, AZ 86002-1149}
\altaffiltext{4}{The University of Arizona, Steward Observatory, Tucson, AZ 8572
1}
\altaffiltext{5}{Spitzer Science Center, California Institute of Technology, 1200 East California Boulevard, Pasadena, CA 91125}
\altaffiltext{6}{Max-Planck-Institut f\"ur Extraterrestrische Physik, Geissenbachstr. 1, D-85741 Garching, Germany}
\altaffiltext{7}{Apache Point Observatory, P. O. Box 59, Sunspot, NM 88349-0059}

\begin{abstract}
{\em XMM-Newton} observations of the Polar SDSSJ155331.12+551614.5 reveal that all the
X-ray flux emerges at energies $<$2 keV. The best fit to the spectrum is with
a thermal plasma with kT=0.8 keV plus a 20-90 eV black body, yielding a thermal
X-ray luminosity of 8.0-9.5$\times$10$^{28}$ ergs s$^{-1}$.
The low temperature and X-ray luminosity, together with the lack of variation
of the X-ray flux during the observations, are all consistent with an
extremely low accretion rate that puts the system 
in the bombardment regime of accretion, rather than accretion involving
 a standoff shock. 
It is likely that the observed X-rays 
originate from the M dwarf secondary star, thus providing a base  
activity level for late main sequence stars in close binaries.
SDSSJ132411.57+032050.5 is detected by {\em XMM-Newton} at the faint EPIC-pn 
count rate of 0.0012$\pm$0.0003
giving an upper limit to the X-ray luminosity of $\sim$7$\times$10$^{28}$ ergs 
s$^{-1}$ for a distance of 300 pc, which is also consistent with the
above scenario.
\end{abstract}


\section{Introduction}

The early results from the Sloan Digital Sky Survey (SDSS; York et al. 2000,
Abazajian et al. 2003, 2004) have shown
that this survey is very effective in finding the shortest period (oldest)
and lowest mass transfer rate (faintest) cataclysmic variables (Szkody
et al. 2002; 2003a). Included among the new discoveries
 are two systems containing magnetic white dwarfs (i.e. Polars) with 
extreme cyclotron features; that is, very narrow, highly polarized cyclotron emission
harmonics that show dramatic changes during the orbital cycle (Szkody et al 2003b). The lack of
strong Balmer emission lines and the narrow appearance of the cyclotron features
imply very low temperature plasmas ($<$1 keV) and accretion rates of 
$<$10$^{-13}$M$_{\odot}$ yr$^{-1}$ (Ferrario et al. 2002). In this regime, the 
energy of the incoming
flow is dissipated through 
particle collisions in the white dwarf atmosphere rather than in an accretion 
shock (Wickramasinghe \&
Ferrario 2000), a situation which is sometimes termed a ``bombardment solution''
(Kuijpers \& Pringle 1982).

In order to check the consistency of this model, we obtained X-ray measurements
of these two systems SDSSJ155331.12+551614.5 and SDSSJ132411.57+032050.5 
(hereafter abbreviated as 
SDSSJ1553 and SDSSJ1324 for simplicity) with the {\em XMM-Newton} satellite. From the data
shown in Szkody et al. (2003b), these objects are known to have orbital 
periods of 4.39 hr and 2.6 hr respectively, derived from the $>$1 mag
sinusoidal modulations in the V and R
light
curves created by the changing cyclotron harmonics. The sinusoidal
shapes of the variations 
imply that the magnetic poles are always in view. 
In each system, the wavelengths of the observed cyclotron harmonics are consistent with
the 3rd and 4th harmonics in a magnetic
field near 60 MG. The presence and strength 
of TiO bands in SDSSJ1553 imply a secondary star of spectral type near M5V,
and assuming a main-sequence mass-radius relation, 
a corresponding distance of 100 pc. The same assumptions for an M6V star in
SDSSJ1324 yield a
distance of $\sim$300-500 pc. 

Detailed modelling of the
spectrum of SDSSJ1553 (Ferrario et al. 2002) indicates that it has an
accretion rate three orders of
magnitude smaller than a typical Polar system, 
and an underlying very cool white dwarf with
temperature of only 5000-8000K. 
With these very unusually low accretion rates and cool white dwarfs, X-rays
are not expected from the accretion spots (Woelk \& Beuermann 1992). 
The ROSAT All Sky Survey
provides upper limits of 0.04 and 0.02 c s$^{-1}$ for SDSSJ1553 and SDSSJ1324.

A recent survey of 37 Polars using {\em XMM-Newton} showed that 16 were in a
low or reduced accretion state (Ramsay et al. 2004), with 6 not detected.
Of the 10 with detections, 8 showed orbital variations, indicating that
some accretion was still occurring. Of the 2 systems with no
orbital variation (CP Tuc and MN Hya), the observation times were too short
to provide a spectrum. However, by assuming a 1 keV thermal plasma and an
absorption of 1$\times$10$^{20}$ cm$^{-2}$, Ramsay et al. (2004) inferred   
X-ray luminosities for the two systems that were 15-20 times 
greater than that expected for normal late main sequence stars.
With our long observation times on the extremely low accretion systems 
SDSSJ1553 and SDSSJ1324,
we are able to accomplish a spectral fit for SDSSJ1553 and obtain 
better luminosity 
constraints for the low states in both Polars.

\section{Observations}

The {\em XMM-Newton} satellite 
observes with three X-ray cameras
(the EPIC-pn has about twice the sensitivity of either the EPIC-MOS1
or MOS2; Turner et al. 2001) as well as X-ray reflection grating
spectrographs (our two systems were too faint for
any useful data from the spectrographs) and an optical monitor (OM; Mason et al. 2001). The OM was
used with a V filter for SDSSJ1553 and a white light filter for SDSSJ1324. 
The data were analyzed following the ABC guide from the {\em XMM-Newton} US GOF
\footnote{http://heasarc.gsfc.nasa.gov/docs/xmm/abc/abc.html} and 
analysis threads from the main Vilspa\footnote{
http://xmm.vilspa.esa.es/external/xmm$_{-}$sw$_{-}$cal/sas.shtml} websites,
using the SAS v6.0 and calibration files current to 2004 March 23. This included reprocessing of ODF files to produce new event files.  These
  were then screened using the standard expressions, and energies were restricted to the range 0.1-10 keV.  Finally, revised good time
  intervals were derived from high energy background light curves to exclude 
intervals of high background flaring.
For the pn, source data were extracted using a 320 pixel radius circular 
aperture (this only encircles $\sim70\%$ of the energy, but reduces the 
background
contribution for this
  low count-rate source) 
and background was selected from adjacent rectangular regions at comparable 
detector Y locations as the target.  As advised in the guide, a conservative
  choice of event selection, pattern = 0, was adopted.  For the two MOS detectors,
  a similar aperture size was used, but the background was determined from an 
annulus surrounding the source, chosen to exclude a neighboring
  faint source, and an event pattern$\leq$12 was also applied.

Light curves for both source and background were extracted using the SAS task 
{\tt evselect}, employing the same extraction regions as described above, but 
with a
less conservative pattern$\leq$4 for the pn events.  The task {\tt lccor} scaled
 and applied background subtraction, dead-time and vignetting
corrections, to produce net source light curves with 100s binning.  Finally, the
 time stamps were converted from JD(TT) to HJD(UT), using
FTOOLS\footnote{http://heasarc.gsfc.nasa.gov/lheasoft/ftools/} tasks. 

The X-ray observation of SDSSJ1553  began
 on 2003 February 17 but was
halted prematurely due to high radiation levels. Additional data were obtained
on 2003 March 12 to complete the observation.
The OM 
provided useful data during the X-ray observations (Figure 1),
while higher time resolution, ground-based CCD differential 
photometry in V and open
 filters was obtained on 2003 February 19 and 22 and
June 29 using the
United States Naval Observatory (USNO) 1m telescope (Figure 2). 
The light
curves all show the same variations as present in Szkody et al 
(2003b),
indicating that the accretion state was similarly low. The variation in the V
and open filters is caused by the changing strength of the n=3 cyclotron
harmonic at 6200\AA.  In addition,
the ground-based and OM optical light curves allow a relative phasing of the X-ray data. 
Due to the high radiation levels and other screening, the good time intervals
of the X-ray data only span portions of the total observation length.
With phase
0 defined as the minimum of the optical variation, the February
good time intervals cover phases 0.75-1.43 (87.61-87.73 in Figure 1) and the 
March observation covers phases 
0.88-1.49 (111.24-111.35 in Figure 1). Thus, most of the orbit is covered,
and the portions from minimum to maximum light are sampled in each
X-ray observation.

SDSSJ1324 was observed with {\em XMM-Newton} on 2004 Jan 25. The source was 
barely detected 
with
the most sensitive EPIC-pn instrument, so no time resolution was possible.
The OM data with 10 min bins and the USNO data from the night before
the {\em XMM-Newton} observations (Figure 3) show the 2.6 hr variation
previously identified as the orbital period.
The X-ray observations of both systems are summarized in Table 1.

\begin{figure}[!tb]
\resizebox{.48\textwidth}{!}{\rotatebox{0}{\plotone{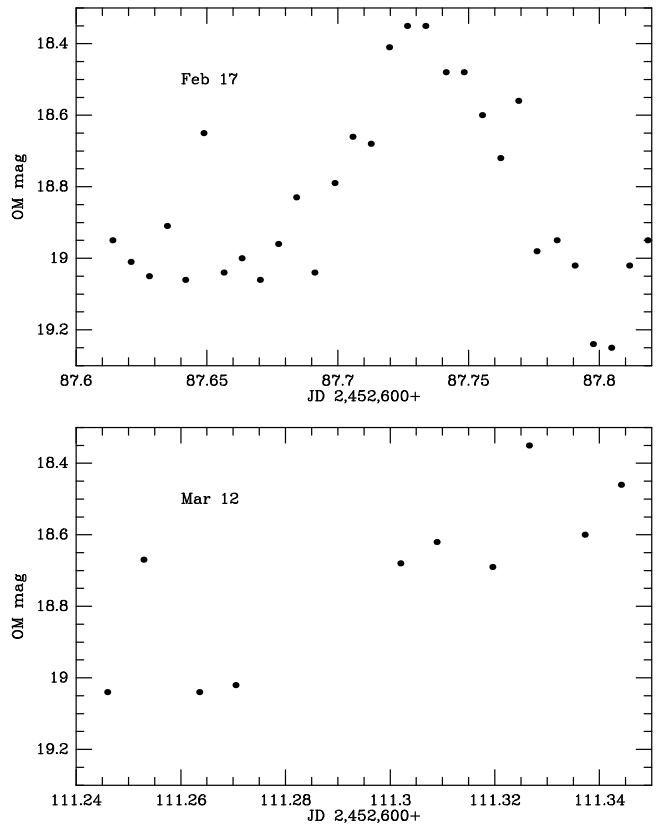}}}
\caption{OM $\it{V}$ light curves of SDSS~J1553 during the February and March {\em XMM-Newton} observations. Data are binned into 10
minute segments with individual 1$\sigma$ errors of 0.06 mag.}
\end{figure}

\section{Results}

\subsection{SDSSJ1553}

The count rates for the February and March datasets on SDSS1553 are consistent
with each other, and the binned X-ray light curves show no variation throughout
the observation intervals. 
A plot of X-ray count rate vs energy (Figure 4) for the EPIC-pn, -MOS1 and -MOS2 data 
shows that all of the observed X-ray
flux for SDSSJ1553 emerges below 2 keV. 
A joint fit was performed in XSPEC\footnote{http://heasarc.gsfc.nasa.gov/docs/xanadu/xspec/index.html} 
on all the pn and MOS data from each of the two runs.  We binned the data to 
give $>5$ counts per bin and used
Cash statistics to find the best overall fit. To accommodate any discrepancies 
in the relative pn and MOS effective area calibrations, the
normalizations were allowed to vary. Similarly, the normalizations between the 
February and March runs were allowed to vary.
SDSSJ1553 is nearby ($d=100$ pc) and at a Galactic latitude of 47 deg, so 
the column density to the source is expected to be low. Hence, this parameter
was fixed at 1.3$\times$10$^{20}$ cm$^{-2}$, the value
obtained from the HEASARC column density tool\footnote{http://heasarc.gsfc.nasa.gov/cgi-bin/Tools/w3nh/w3nh.pl}.

\begin{figure*}[!tb]
\resizebox{0.9\textwidth}{!}{\rotatebox{0}{\plotone{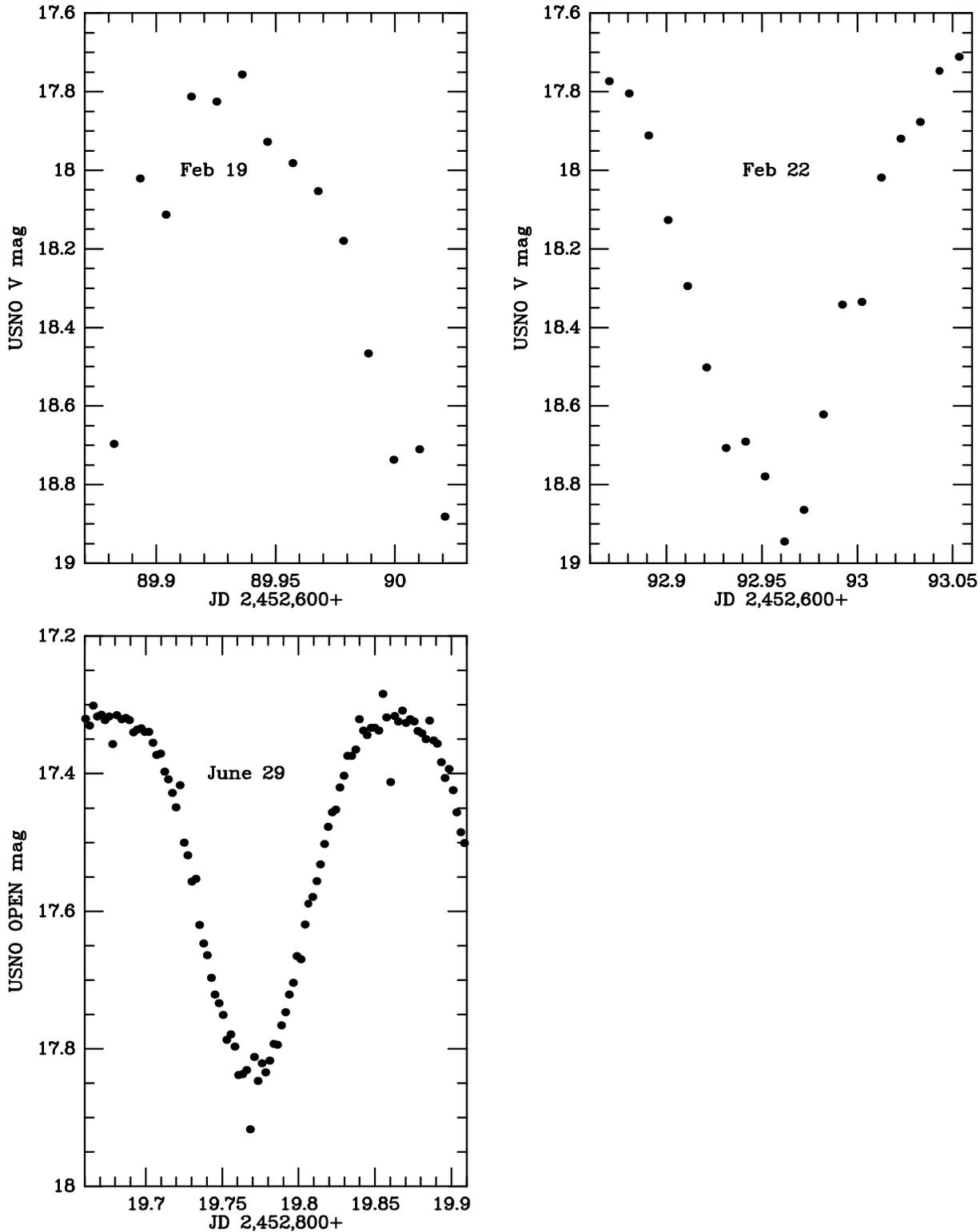}}}
\caption{Ground-based optical light curves of SDSS~J1553 on February 19 and 22 and June 29, which show the same orbital modulations as apparent in Szkody et al. (2003a). February 19 data have
1$\sigma$ error bars of 0.08 mag, February 22 of 0.05 mag, and June 29 of 0.01 mag.}
\end{figure*}

\begin{deluxetable*}{lcccllllll}
\tablewidth{0pt}
\tablecaption{Summary of {\em XMM-Newton} Observations}
\tablehead{
\colhead{SDSSJ} & \colhead{Date} & \colhead{UT} & \colhead{PN(c/s)} &
\colhead{PN(s)} & \colhead{MOS1(c/s)} & \colhead{MOS1(s)} & \colhead{MOS2(c/s)} &
\colhead{MOS2(s)} & \colhead{OM(s)} }
\startdata
1553 & 2003-02-17 & 02:29-07:38 & 0.020$\pm$0.002 & 5130 & 0.009$\pm$0.001 &
6396s & 0.006$\pm$0.001 & 6394s & 16700 \nl
1553 & 2003-03-12 & 17:39-20:23 & 0.026$\pm$0.003 & 7018 & 0.010$\pm$0.001 &
9085 & 0.007$\pm$0.001 & 9192 & 6000 \nl
1324 & 2004-01-25 & 02:27-07:28 & 0.0012$\pm$0.0003 & 16234 & &  & & & 13824 \nl
\enddata
\end{deluxetable*}

Because of the uncertainties in the calibration at the lowest energies, fits
were attempted using energies $>$0.15 keV for pn and $>$0.2 keV for MOS, as well
as more conservative limits of $>$0.3 keV (pn) and $>$0.5 keV (MOS).
A thermal plasma (mekal) model with kT=0.78$\pm$0.03 keV 
provided the best fit in both cases.  However, Figure 4 shows that there may
be an additional soft component present at energies $<$0.3 keV. This can
be fit with a black body of roughly 46 eV which is poorly constrained
(20-90eV is possible).
Table 2 gives the 
parameters of this type of fit and it is shown in Figure 4.  We note that there
 are larger residuals to the fit at around 0.8 keV, most likely
due to a combination of the noise and low energy resolution of the data and the 
complexity of the (unresolved) emission lines in this region (see  Mukai et
et al. (2003) for higher resolution X-ray spectra of cataclysmic variables). 
The unabsorbed total flux (0.2-10 keV) for the mekal fit alone is 6.7-7.9$\times$10$^{-14}$ ergs cm$^{-2}$ s$^{-1}$. Using the distance determined
from the observed M star (100 pc) yields an X-ray luminosity of 8.0-9.5$\times$10$^{28}$ ergs s$^{-1}$.

\begin{figure}[!tb]
\resizebox{.48\textwidth}{!}{\rotatebox{0}{\plotone{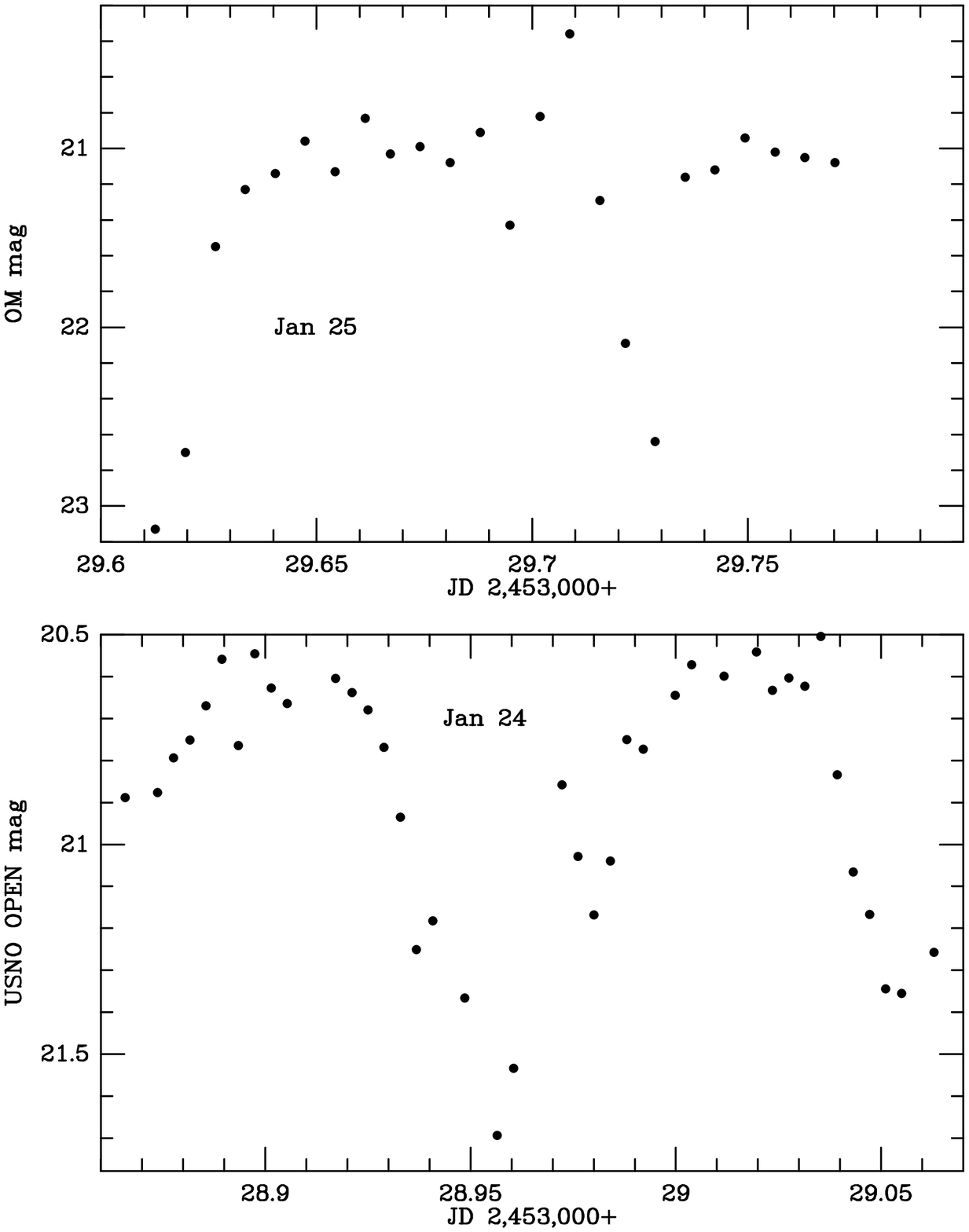}}}
\caption{OM white light curve of SDSS~J1324 binned by 10 minutes during the January {\em XMM-Newton} observation (top) and white light curve from USNO the night before the X-ray observation (bottom). USNO data points have 1$\sigma$ errors ranging from 0.07 at 20.5 mag to 0.15 at 21.5 mag (except for data at times 28.96-28.97, which have errors up to 1 mag because of passing clouds).}
\end{figure}

The lack of hard X-rays and absence of modulation in the light curve  
corroborate that the X-ray source is not associated with an accretion shock at
the magnetic pole, and that this low accretion rate is within the
bombardment regime. However, the origin of X-ray flux that emerges even at these
low rates is unclear.  Ramsay et al. (2004) concluded that
there was still some level of ongoing accretion even in the systems showing no
orbital modulation. However, this conclusion was based on assumptions
about temperature and absorption, and a lack of information about
the exact accretion level of CP Tuc and MN Hya. 
The pn 
count rates for CP Tuc and MN Hya are comparable to SDSSJ1553, even though 
the former are 2--5 times further away. This alone implies that they have a higher 
level of accretion than SDSSJ1553.

\begin{figure}[!tb]
\resizebox{.48\textwidth}{!}{\rotatebox{270}{\plotone{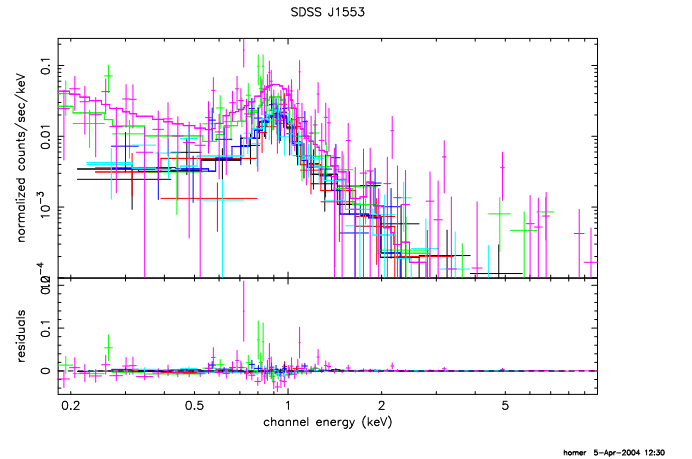}}}
\caption{{\em XMM-Newton}/EPIC pn and MOS data for both February and March data sets with best-fit thermal MEKAL+blackbody model. The discrepancies around 0.8 keV simply reflect the combination of the low spectral resolution and the complex of emission lines in the emitting plasma in this region.}
\end{figure}

Many Polars display a
variety of low states, corresponding to different levels of accretion. The prototype,
AM Her, is one such example, exhibiting high states with V mag of 12.5, 
intermediate states near 13.5 and low states near 15.2. Fabbiano (1982)
observed AM Her at an intermediate state of 14.5 and found the usual high state 
components (13.5 keV from the shock and 30 eV from the
heated white dwarf) replaced by a single 9 keV component. De Martino et al. (1998)
observed AM Her in one of its lowest states (V=15.1) and found thermal X-ray
flux with a temperature near 5.8 keV for part of the observation and then a
drop in X-ray flux by a factor of 7 to the lowest level yet seen. This lowest
portion showed a temperature $>$3.6 keV, with a corresponding luminosity of
3$\times$10$^{30}$ ergs s$^{-1}$ and an accretion rate of 
$\ge$4$\times$10$^{-13}$M$_{\odot}$/yr. Since this temperature is higher than
expected for coronal emission (which is usually in the 0.25-1.7 keV range), they 
concluded
that there was likely still some residual accretion, but that the secondary
could have a substantial contribution to the X-ray flux.  

SDSSJ1553 has
the lowest accretion level known (3 orders of magnitude below most Polars and
an order of magnitude below AM Her in its lowest state), 
a value that is consistent with wind accretion
from the companion. There is no broad line emission, nor evidence of any type,
 for
a mass transfer stream
from the secondary to the white dwarf. Thus, this system might best serve as
a template for the base level of X-ray flux in a Polar. 

The extremely low white dwarf temperature of SDSSJ1553 indicates that the
average accretion rate over about the last million years (the Kelvin-Helmholtz 
time scale
of the compressionally heated envelope) cannot have been larger than 
$\sim$3$\times$10$^{-12}$M$_{\odot}$ yr$^{-1}$
(Townsley \& Bildsten 2004).  This raises the distinct possibility that
the secondary in SDSSJ1553 has not yet contacted its Roche lobe, a fact
consistent with the inferred M5 spectral type if the main-sequence mass-radius
relation applies.  As such, SDSSJ1553 would be the first pre-Polar known, and
possibly a precursor to objects such as RXJ1313.2-3259 (G\"ansicke et al. 2000).

The temperature of the plasma that is emitting X-rays in SDSSJ1553 is in the regime
for a coronal source.
For its orbital period and spectral type, the mass
of the secondary is likely near 0.4M$_{\odot}$ (Warner 1995). From the mass-luminosity
relation of Malkov, Piskunov \& Shipil'kina (1997), the total luminosity of
the secondary should be near 1$\times$10$^{32}$ ergs s$^{-1}$. Given
that the secondary
 should be in the saturated regime with L$_{x}$/L$_{bol}$ $<$ 10$^{-3}$,
the X-ray luminosity should be less than 10$^{29}$ ergs s$^{-1}$. The 
observed X-ray 
luminosity is compatible with this limit. 
Thus, SDSSJ1553 appears to be an excellent candidate for
coronal emission of X-rays. 

Even though the infalling gas cools almost exclusively through electron
cyclotron emission at these very low accretion rates, a blackbody component is
expected due to surface heating from the downward-directed cyclotron flux. The
``bombardment'' calculations of Woelk \& Beuermann (1992, 1993) indicate that
irradiation occurs from a Rosseland optical 
depth $\tau \approx 10^{-3}$ to $10^{-4}$
and that reprocessing from the underlying photosphere should appear as an EUV
component with a temperature of $\sim$10$^5$~K.  The spectral fit with a
mekal+blackbody model gives a total 0.01$-$10~keV flux of 1.2$-$1.5$\times10^{-13}$
ergs cm$^{-2}$ s$^{-1}$, with the blackbody contributing 32$-$35\% of this
flux.  This is $\sim$25\% of the observed cyclotron flux (Ferrario et al.
2002), and for a distance of 100 pc amounts to a luminosity of
4.7$-$6.2$\times10^{28}$ ergs s$^{-1}$.  Our temperature estimate of 50~eV
(600,000~K) leads to an unrealistically small
 radius for the emission region of 5$\times10^4$~cm, or
0.006\% of the radius of a 0.7~$M_{\odot}$ white dwarf. Unfortunately, the
upturn in the flux occurs at energies where the calibrations are most uncertain,
so the temperature is very uncertain and the mere existence of this component
must be confirmed (e.g. with more sensitive/better calibrated X-ray
observations).  Ramsay et al. (2004) used the UV filters on their OM
observations to determine that a low temperature component was present outside
the bandpass of the XMM detectors.  But since their objects exhibited higher
accretion rates and X-ray luminosities, it is difficult to directly compare
results.

\begin{deluxetable}{lccccc}
\tablewidth{0pt}
\tablecaption{Spectral Fits}
\tablehead{
\colhead{SDSSJ} & \colhead{Model}
& \colhead{kT$_{mek}$(keV)} & \colhead{N$_{H}$\tablenotemark{a}} &
 \colhead{kT$_{bb}$(eV)} &
\colhead{F$_{unabs}$\tablenotemark{a}}}
\startdata
1553 & Mekal & 0.78$\pm$0.026 & 1.3 & &  6.7-7.9 \nl
1553 & Mekal+BB & 0.79$\pm$0.026 & 1.3  & 46 & 12-15 \nl
1324 & Mekal & 0.2 & 2.0  & & $<$0.6 \nl
\enddata
\tablenotetext{a}{Units for N$_{H}$ are 10$^{20}$ cm$^{-2}$. F is unabsorbed
flux over energies of 0.2-10 keV for mekal and 0.01-10 keV for mekal+BB.
Flux units are $\times$10$^{-14}$ ergs cm$^{-2}$ s$^{-1}$.}
\end{deluxetable}

\subsection{SDSSJ1324}

SDSSJ1324 was too faint to provide more than an upper limit to the
X-ray flux. In the pn, the background contribution and faint source counts
are comparable within our analysis aperture.
A very crude ``spectrum'' was constructed from 4 energy bins, and a background
power law model was scaled to contribute 50\% in the source aperture and
then added to the source model. The column found from the HEASARC tool 
(2$\times$10$^{20}$ cm$^{-2}$) was then fixed and the temperature was stepped 
from 0.2-1 keV.
The best fit was at 0.3 keV, but for an upper limit we used 0.2 keV to
determine a 0.2-10 keV flux of 4$\times$10$^{-15}$ ergs cm$^{-2}$ s$^{-1}$ and
an unabsorbed flux of 6$\times$10$^{-15}$ ergs cm$^{-2}$ s$^{-1}$. The
corresponding X-ray luminosity is 7$\times$10$^{28}$ ergs s$^{-1}$ for an
assumed distance of 300 pc.

\section{Conclusions}

Our X-ray observations of the extremely low accretion rate Polar SDSSJ1553
reveal a low count rate of 0.030$\pm$0.003 c/s, no detectable 
variation of the X-ray flux throughout
the orbit (while the optical shows large variations due to the changing 
amplitude of the cyclotron harmonics),  and a spectrum that can be modeled by 
a thermal plasma
with a temperature near 0.8 keV. The X-ray flux translates into a luminosity
below 10$^{29}$ ergs s$^{-1}$. The low count rate of SDSSJ1324 
(0.0012$\pm$0.0003) is also consistent with this X-ray luminosity. 
All these characteristics imply an 
origin of the X-rays from coronal emission on the secondary star,
rather than an accretion shock. The sensitivity of $\it{XMM-Newton}$ has allowed
the base level of X-ray emission from a close binary to be measured. 
Further X-ray monitoring of these stars whose mass transfer has turned off can
reveal if the flare activity of M stars in close binaries is enhanced due
to the close proximity of its companion white dwarf.

\acknowledgments

We acknowledge an anonymous referee for helpful suggestions on the evolutionary
state.
This work was funded by NASA XMM grants NAG5-12938 and NNG04GG66G and NSF grant AST-02-05875 to the
University of Washington. This work is based on observations obtained with
{\em XMM-Newton}, an ESA science mission with instruments and contributions
directly funded by ESA member statess and the USA (NASA).

Funding for the creation and distribution of the SDSS Archive has been provided by the Alfred P. Sloan Foundation, the Participating Institutions, the National Aeronautics and Space Administration, the National Science Foundation, the U.S. Department of Energy, the Japanese Monbukagakusho, and the Max Planck Society. The SDSS Web site is http://www.sdss.org/.
The SDSS is managed by the Astrophysical Research Consortium (ARC) for the Participating Institutions. The Participating Institutions are The University of Chicago, Fermilab, the Institute for Advanced Study, the Japan Participation Group, The Johns Hopkins University, Los Alamos National Laboratory, the Max-Planck-Institute for Astronomy (MPIA), the Max-Planck-Institute for Astrophysics (MPA), New Mexico State University, University of Pittsburgh, Princeton University, the United States Naval Observatory, and the University of Washington.

%\clearpage

%\tiny
%\normalsize
%\clearpage
%\plotone{szkody.fig1.ps}
%\clearpage
%\plotone{szkody.fig2.ps}
%\clearpage
%\plotone{szkody.fig3.ps}
%\clearpage
%\psfig{figure=szkody.fig4.ps,width=6in,angle=270}
\end{document}